\documentclass[]{iopart}
\usepackage{iopams}
\usepackage{graphicx}
\include{Josephson.bib}
\begin{document}

\title[]{Josephson Junctions as Threshold Detectors of the Full Counting Statistics: Open issues}%
\author{Tom\'{a}\v{s} Novotn\'{y}}%
\address{Department of Condensed Matter Physics, Faculty of
Mathematics and Physics, Charles University, Ke Karlovu 5, 121 16
Prague, Czech Republic}%
\ead{tno@karlov.mff.cuni.cz}%

\begin{abstract}
I study the dynamics of a Josephson junction serving as a threshold
detector of fluctuations which is subjected to a general
non-equilibrium electronic noise source whose characteristics is to
be determined by the junction. This experimental setup has been
proposed several years ago as a prospective scheme for determining
the Full Counting Statistics of the electronic noise source. Despite
of intensive theoretical as well as experimental research in this
direction the promise has not been quite fulfilled yet and I will
discuss what are the unsolved issues. First, I review a general
theory for the calculation of the exponential part of the
non-equilibrium switching rates of the junction and compare its
predictions with previous results found in different limiting cases
by several authors. I identify several possible weak points in the
previous studies and I report a new analytical result for the linear
correction to the rate due to the third cumulant of a non-Gaussian
noise source in the limit of a very weak junction damping. The
various analytical predictions are then compared with the results of
the developed numerical method. Finally, I analyze the status of the
so-far publicly available experimental data with respect to the
theoretical predictions and discuss briefly the suitability of the
present experimental schemes in view of their potential to measure
the whole FCS of non-Gaussian noise sources as well as their
relation to the available theories.
\end{abstract}
\section{Introduction}

Josephson junctions (JJs) were proposed as threshold detectors of
the Full Counting Statistics (FCS) by Tobiska and Nazarov
\cite{Tobiska2004} and independently by Pekola \cite{Pekola2004} in
2004. Since then there has been continuing effort to implement the
proposed schemes experimentally as well as to improve them and
better understand their potential theoretically. The original scheme
by Tobiska and Nazarov \cite{Tobiska2004} proposed using overdamped
JJ as the threshold detector. This appears to be problematic since
in the overdamped junction when the effective phase particle
overcomes the tilted washboard potential barrier it gets immediately
retrapped in the adjacent minimum. This results in the phase
diffusion which, however, does not yield enough sensitivity for
detecting the whole FCS. This could in principle be overcome by
employing a negative-inductance device which apparently hasn't
appealed to the experimentalists enough to actually implement it.
Instead they opted for an obvious alternative to use underdamped
junctions where, under suitable conditions, once the particle
overcomes the first barrier it keeps on sliding down the potential
thus producing finite voltage. Thus, the switching of an underdamped
junction between the supercurrent (static phase) and running (finite
phase velocity, i.e.~finite voltage) state would provide a prime
example of a threshold detector. Unfortunately, this
innocently-looking change in the setup dramatically changes the
level of difficulties involved in the theoretical analysis. This
paper addresses those difficulties in some detail.

The structure of the paper is the following. In the next section
\ref{Sec2} I report the theoretical concept of calculating the
non-equilibrium escape rate due to a non-Gaussian noise source whose
FCS is to be determined. The general theory based on the WKB-like
approximation for the weak noise intensity is further carried out to
an analytical result in case of the linear perturbation theory in
the third cumulant for very weak junction damping in subsection
\ref{subsec21}. In this subsection I also make comparison with
alternative existing theories. In the next subsection \ref{subsec22}
the full theory is numerically implemented and the numerical results
in an experimentally relevant regime are discussed and further
compared with various analytical predictions. In section \ref{Sec3}
I briefly raise some experimentally relevant questions such as what
is the effect on the rate asymmetry of the nominally subleading
terms entering the rate and whether one can actually experimentally
leave the linear regime and achieve the measurement of the whole FCS
of a noise source. In the last section \ref{Concl} I summarize what
has been achieved in this work and review the remaining open
problems.

\section{Theoretical calculation of the non-equilibrium escape
rate} \label{Sec2}

The Josephson element in an electrical circuit is often modeled as a
current biased ($I_b$) resistively ($R$) and capacitively ($C$)
shunted ideal JJ with the Josephson current-phase relation
$I_J(\varphi)=I_0\sin\varphi$. The voltage across the junction is
determined by the second Josephson relation
$V_J=\dot{\varphi}\,\hbar/2e$ with the time-derivative denoted by
the dot. Moreover, due to the action of ubiquitous thermal
(Gaussian) noise $\xi(t)$ characterized by the temperature $T$ and
non-equilibrium electronic noise $\eta(t)$ from the measured device
whose FCS is to be determined, the JJ is subjected to stochastic
forces and its dynamics is thus described by the following Langevin
equation (RCSJ model)
\begin{equation}\label{Langevin_eq}
    \ddot{\varphi}+\frac{1}{RC}\dot{\varphi}+\frac{2e}{C\hbar}(I_0\sin\varphi-I_b)=\xi(t)+\eta(t).
\end{equation}
In a realistic experimental situation the current-bias assumption
can be inadequate and one may need to generalize the above model.
The general consequences of an imperfect current bias are so called
environmental or ``cascade" corrections to the measured cumulants of
the source FCS which were studied in previous works
\cite{Sukhorukov2007,Grabert2008}. They could be straightforwardly
included here in the same spirit as in those works, especially
Ref.~\cite{Grabert2008}, but since they appear to be of minor
importance in the so far reported experiments I will neglect them.

In this study I will consider in detail exclusively the simplest
case of the Poissonian shot noise $\eta(t)$ corresponding to the
measured device being a tunnel junction. In such a case $\eta(t)$ is
just a train of $\delta$-function-like spikes which are separated by
an exponentially distributed waiting time with a single parameter
(mean waiting time) being the mean (particle) current $I_m/e$
flowing through the tunnel junction. This case is also the only one
studied experimentally by this type experiments to date. Assuming
the temporal width of the pulses composing $\eta(t)$ to be very
small compared to a characteristic time of the junction dynamics
(which is its plasma frequency $\omega_{p0}=\sqrt{2eI_0/\hbar C}$)
one can obtain a master equation (analogous to the Fokker-Planck
equation in case of Gaussian noise only) for the probability density
$W(x,v,t)$ in dimensionless units $t\omega_p\rightarrow t,\,
\varphi\rightarrow x,\,\dot{\varphi}/\omega_{p0}\rightarrow v$
\begin{eqnarray}\label{FPeq}
  \frac{\partial W}{\partial t} &= &-v\frac{\partial W}{\partial x}+Q^{-1}\frac{\partial (vW)}{\partial
v}+\!\!\left(\sin x-s+\frac{I_m}{I_0}\right)\frac{\partial
W}{\partial v} +Q^{-1}\frac{k_BT}{E_J}\frac{\partial^2W}{\partial
v^2} \nonumber\\
    & +&\frac{I_m}{I_0\lambda}\left[\exp\left(-\lambda\frac{\partial}{\partial
v}\right)-1\right]W,
\end{eqnarray}
with $s=I_b/I_0$ the rescaled bias current, $Q=RC\omega_{p0}$ the
quality factor of the (unbiased \footnote{Note that my definition of
the quality factor differs from that in Ref.~\cite{Ankerhold2007}
where a bias-specific quality factor is used instead.}) junction and
$\lambda=\sqrt{e^2/C E_J}$ with the Josephson energy of the junction
proportional to the critical current $E_J=I_0\hbar/2e$. The last
term in the equation can be identified as stemming from the cumulant
generating function of the Poissonian process $F_{\rm
Poisson}(x)=I_m(\exp x-1)$ \footnote{Here I assume $I_m\geq 0$ and
the Eq.~(\ref{FPeq}) holds for the positive polarity of the
tunneling current. The opposite polarity would just change the sign
in the exponential in Eq.~(\ref{FPeq}).} which suggests how to deal
with non-Poissonian noise source provided the Markovian
approximation is made. Thus, the substitution
$I_m\left[\exp\left(-\lambda\partial/\partial
v\right)-1\right]\rightarrow F(-\lambda\partial/\partial v)$ for
general noise sources described by the cumulant generating function
$F(x)$ generalizes the particular results shown here for the
Poissonian process to arbitrary noise sources as long as the
Markovian approximation is justified
\cite{Tobiska2004,Ankerhold2007}.

In order to calculate the escape rates of the junction from the
supercurrent branch (zero voltage state with a static phase
$\varphi$) to the running state (finite voltage across the junction
with non-zero phase velocity $\dot{\varphi}$) in the low noise limit
we use the standard technique known as the singular perturbation
theory in the mathematical literature \cite{Kevorkian1996} or as WKB
method in the physical context \cite{Hanggi1990, Risken1989}. It
consists in making the ansatz $W(x,v,t)=\exp[S(x,v,t)/\theta]$ for
the probability density $W(x,v,t)$ with $\theta$ being a small
parameter related to the noise intensity: $\theta\equiv k_B T_{\rm
eff}/E_J\equiv k_B T/E_J+Q\lambda I_m/2I_0=k_B T/E_J+ e R I_m/2E_J$.
Thus, $\theta$ is a dimensionless effective temperature of the
junction due to the summed effect of the thermal noise and the
Gaussian part of the non-equilibrium noise
\cite{Tobiska2004,Sukhorukov2007,Ankerhold2007,Pekola2005,Huard2007}.
When this ansatz is put into Eq.~(\ref{FPeq}) and the lowest order
in $\theta$ is only retained (corresponding to the WKB approximation
and justified for small $\theta\ll1$) we obtain the following
Hamilton-Jacobi (HJ) equation, i.e.\ a first order partial
differential equation for $S(x,v,t)$
\begin{eqnarray}\label{HJeq}
\frac{\partial S}{\partial t}&=&-v\frac{\partial S}{\partial
x}+(\sin x-s)\frac{\partial S}{\partial v}+Q^{-1}v\frac{\partial
S}{\partial v}+Q^{-1}\left(\frac{\partial S}{\partial
v}\right)^2+\frac{I_m}{I_0}\sum_{n=3}^{\infty}\frac{1}{n!}\left(\frac{\lambda}{\theta}\right)^{n-1}\left(-\frac{\partial
S}{\partial v}\right)^n\nonumber\\
&=&-v\frac{\partial S}{\partial x}+(\sin x-s)\frac{\partial
S}{\partial v}+Q^{-1}v\frac{\partial S}{\partial
v}+Q^{-1}\left(\frac{\partial S}{\partial
v}\right)^2+\frac{\theta}{I_0\lambda}\tilde{F}_{\rm
Poisson}\left(-\frac{\lambda}{\theta}\frac{\partial S}{\partial
v}\right).
\end{eqnarray}
For a general noise source the last term (given by the sum) in the
preceding equations would be replaced by the corresponding
expression $\tilde{F}(x)=F(x)-F'(0)x-F''(0)x^2/2$, i.e. by the
reduced cumulant generating function with the first two moments
(mean current and the zero-frequency noise) subtracted (notice that
$F(0)=0$ by definition).

This Hamilton-Jacobi equation can be solved via the method of
characteristics, i.e. one can recast the equation as a dynamical
system in a 4-dimensional phase-space $[x(t),v(t),p(t)\equiv\partial
S/\partial x,y(t)\equiv\partial S/\partial v]$ governed by the
auxiliary Hamiltonian
\begin{equation}\label{Hamiltonian}
    \mathcal{H}=vp-(\sin
    x-s)y-Q^{-1}y(v+y)+\frac{\theta}{I_0\lambda}\tilde{F}\left(-\frac{\lambda}{\theta}y\right).
\end{equation}
The coordinates $x,v$ and their conjugated momenta $p,y$ are then
evolving according to the following equations of motion
\begin{eqnarray}\label{EOM}
\dot{x}&=&\frac{\partial\mathcal{H}}{\partial p}=v,\
\dot{p}=-\frac{\partial\mathcal{H}}{\partial x}=y\cos x,\
\dot{y}=-\frac{\partial\mathcal{H}}{\partial v}=-p+Q^{-1}y,\nonumber\\
\dot{v}&=&\frac{\partial\mathcal{H}}{\partial y}=-(\sin
x-s)-Q^{-1}(v+2y)-\frac{1}{I_0}\tilde{F}'\left(-\frac{\lambda}{\theta}y\right).
\end{eqnarray}
The exponential part of the non-equilibrium escape rate is in this
language determined by the difference of the stationary, i.e.\
time-independent, action between the barrier top and the metastable
minimum of the tilted washboard potential, i.e.\ $\theta\log\Gamma
\propto S(x_{\rm max},0)-S(x_{\rm min},0)$ \cite{Hanggi1990} which
can be either determined by the direct solution of Eq.~(\ref{HJeq})
or by finding the action along the trajectory of the system
(\ref{EOM}) connecting in infinite time (corresponding to the
stationary solution and zero auxiliary energy $\mathcal{H}=0$) the
two fix points $[x_{\rm min},0,0,0]$ and $[x_{\rm max},0,0,0]$
\cite{Kamenev2004}. I will demonstrate both methods in the next two
subsections. At this point I would like to stress that the
overdamped analog of the present problem (with $C\to 0$ in
Eq.~(\ref{Langevin_eq}) when the inertial term $\ddot{\varphi}$ can
be neglected) studied in Ref.~\cite{Tobiska2004} is integrable since
analogous equations to (\ref{HJeq}), (\ref{EOM}) are to be solved in
a 2-dimensional phase-space only and with the help of the
stationarity constraint $\mathcal{H}_{\rm overdamped}(x,p)=0$ one
easily finds the action in Eq.~(\ref{HJeq}) by a quadrature for an
arbitrary strength of the non-equilibrium noise. Unfortunately, this
property does not carry over to the underdamped case where the
energy constraint is not sufficient for integrability. Therefore,
the underdamped problem is conceptually far more difficult than the
originally suggested overdamped model.

\subsection{Linear perturbation theory of the rate asymmetry due to a weak third cumulant}
\label{subsec21}

In this subsection I will present a linear perturbation theory of
the rate asymmetry which is an alternative to the similar previous
approaches by a number of authors
\cite{Sukhorukov2007,Grabert2008,Ankerhold2007}. I will use this
limiting case for the illustration of the general method, which will
be fully developed in the next subsection, and, at the same time,
for pointing out possible discrepancies in the previous studies. As
a by-product I will present a new analytical formula for the rate
asymmetry in the very low damping limit $Q\to\infty$.

Following the previous studies we consider the linear correction to
the escape rate due to a weak third cumulant. To this end we
truncate the sum in the stationary version of the HJ equation
(\ref{HJeq}) to the first order, i.e.\ we consider the effects of
the third cumulant $c_3$ only. Further, we formulate the linear
perturbation theory for an arbitrary potential $V(x)$ in which the
effective particle moves
--- the present case is then recovered by the choice of the tilted
washboard potential describing the JJ (in dimensionless units)
$V(x)=-\cos x - s x$. The resulting HJ then reads
\begin{equation}\label{HJ3}
    0 =-v\frac{\partial S}{\partial x}+V'(x)\frac{\partial S}{\partial v}+Q^{-1}v\frac{\partial S}{\partial
v}+Q^{-1}\left(\frac{\partial S}{\partial
v}\right)^2-c_3\left(\frac{\partial S}{\partial v}\right)^3,
\end{equation}
with $c_3=I_m \lambda^2/6 I_0 \theta^2$. We solve this equation in
the linear order in $c_3$ by linearizing the equation. After
inserting $S(x,v)=S_0(x,v)+c_3 S_1(x,v)$ into the equation, using
the knowledge of the zeroth order solution $S_0(x,v)=-v^2/2-V(x)$
corresponding to the Boltzmann factor due to the thermal Gaussian
noise, and keeping only the linear terms in $S_1$ one obtains
\begin{equation}\label{HJlin}
    v^3=v\frac{\partial S_1}{\partial x}+[Q^{-1}v-V'(x)]\frac{\partial S_1}{\partial
    v}\ .
\end{equation}
It is very unlikely that Eq.~(\ref{HJlin}) could be solved
analytically for  general $Q$. Ankerhold \cite{Ankerhold2007} did
find certain solution to the problem of the rate asymmetry for any
$Q$, however, his solution is not a solution of the above equation
(\ref{HJlin}) as I will discuss later on. Indeed, one should not
expect finding an explicit analytical solution to Eq.~(\ref{HJlin})
for arbitrary $Q$ since it is generally known that the action
$S(x,v)$ (or the ``non-equilibrium potential") develops a dense set
of singularities close to the barrier top \cite{Graham1989}. This
does not happen only in the integrable cases which is certainly the
limit $Q\to 0$ corresponding to the one-dimensional spatial
diffusion and, hopefully, also for $Q\to\infty$ describing the
energy diffusion limit, which is effectively one-dimensional again.

Here, I give an analytic expression for the solution $S_1(x,v)$ in
the limit $Q\to\infty$ for general potential $V(x)$, in particular
for the tilted washboard potential without resorting to its cubic
approximation employed in previous works
\cite{Ankerhold2007,Grabert2008}. We look for the solution of
Eq.~(\ref{HJlin}) with $Q\to\infty$ in the form
$S_1(x,v)=\phi_0(x)+\phi_2(x)v^2/2$ and find a closed set of
equations
\begin{eqnarray*}
       \phi_2'(x) & = 2, \\
       \phi_0'(x) & = V'(x) \phi_2(x).
\end{eqnarray*}
The solution $\phi_2(x)=2(x-x_0),\, \phi_0(x)=2\int^x dy
\,yV'(y)-2x_0 V(x)+C$ contains two arbitrary constants $C,x_0$.
Moreover, one can add an arbitrary solution of the homogeneous part
of Eq.~(\ref{HJlin}) to this particular solution. Solutions to the
homogeneous problem are arbitrary (sufficiently smooth and
differentiable) functions of the particle energy $G(v^2/2+V(x))$,
thus, the freedom in the particular solution can be absorbed into
the homogeneous solution since it just represent a linear function
of the energy. The arbitrariness stemming from the mathematical
solution must be fixed by physical requirements. First, all physical
quantities must be ``gauge invariant" meaning that an arbitrary
constant shift in the potential $V(x)\to V(x)+\Delta$ cannot change
the physical observables. Those are changes of $S_1(x,v)$ between
different points in the phase-space, i.e.\ not just $S_1(x,v)$
itself, rather its partial derivatives $\partial S_1/\partial x$ and
$\partial S_1/\partial v$. The conditions to be satisfied are then
$\partial^2 S_1/\partial x\partial\Delta=0$ and $\partial^2
S_1/\partial v\partial\Delta=0$. Both lead to the same equation
$G''(x)=0$ with the linear function solution. This way, we recover
the freedom stemming from the particular solution but the larger
freedom of the homogeneous solution has been removed. The remaining
uncertainty, being basically just the choice of the origin of
integration $x_0$, since the constant $C$ is harmless, is fixed by
the requirement that in the vicinity of the potential minimum, where
the potential can be approximated as harmonic, all the Gaussian
averages (i.e.\
$<\!\!v\!\!>,\,<\!\!x\!\!>,\,<\!\!x^2\!\!>,\,<\!\!xv\!\!>,\,<\!\!v^2\!\!>$)
of the original Fokker-Planck/master equation must stay intact by
the third cumulant. In the harmonic regime, this is a necessary
consequence of the linearity of the underlying Langevin equation.
This condition implies that the origin of integration must be
identical with the potential minimum $x_0=x_{\rm min}$. In total,
one finally has
\begin{equation}
    S_1(x,v)=2\int_{x_{\rm min}}^x dy \,(y-x_{\rm min}) V'(y)+(x-x_{\rm min})v^2 +
    C,
\end{equation}
yielding for the exponential part of the rate asymmetry
$R_{\Gamma}\equiv\Gamma_+/\Gamma_-$ (factor of 2 stands for the sum
of the two equal contributions to the asymmetry from the two
opposite polarities of the measured current)
$R_{\Gamma}(Q\to\infty)=\exp[2 c_3(S_1(x_{\rm max},0)-S_1(x_{\rm
min},0))/\theta] = \exp\left[2 D_1(s) E_J^2 <\!\!<\!I_m^3\!>\!\!>/C
I_0(k_B T_{\rm eff})^3\right]$ with the function $D_1(s)$ introduced
in Ref.~\cite{Sukhorukov2007} reading
\begin{eqnarray}\label{D1}
  D_1(s) & \equiv& \frac{1}{6}[S_1(x_{\rm max},0)-S_1(x_{\rm min},0)]=\frac{1}{3}\int^{x_{\rm max}}_{x_{\rm min}}dx(x-x_{\rm min})V'(x)\nonumber\\
    & =&\frac{1}{3}\int^{\pi-\arcsin s}_{\arcsin s} dx(x-\arcsin s)(\sin
    x-s)\nonumber\\
    &=&\frac{2}{3}\arccos s\left[\sqrt{1-s^2}-s\arccos s\right]
\end{eqnarray}
where the second part applies to the particular case of the tilted
washboard potential.

The value of $D_1(s)$ at zero is $D_1(0)=\pi/3$ in accordance with
Ref.~\cite{Sukhorukov2007} while the asymptotics for large bias is
$D_1(s\to 1)\approx a(1-s)^2$ with $a=8/9$ which is exactly equal to
the result by Ankerhold \cite{Ankerhold2007} but it actually differs
from SJ's numerical finding $a\approx 0.8$ \cite{Sukhorukov2007}
further supported  by an independent study by Grabert
\cite{Grabert2008} with $a\doteq 0.79$. Although the difference is
not severe being on the order of 10\% only and, therefore,
practically most likely irrelevant, from the conceptual point of
view it matters because all the works claim to calculate the same
quantity for the very same model and, thus, the correct result
should be unique. It's not simple to follow and reproduce SJ's
approach but Grabert's method is very transparent and I have fully
recovered his numerical findings. Minor generalization of his
approach to the full tilted washboard potential (Grabert uses the
usual cubic approximation to the tilted-washboard potential for a
large bias $s\to 1$) gives $D_1(0)=\pi/3$ (within numerical
precision) and $a\doteq 0.79$ for $s\to 1$. The discrepancy with the
above analytical result is thus not a problem of the numerical
precision of works \cite{Sukhorukov2007,Grabert2008} but a
conceptual problem. Since Grabert uses the trajectory approach of
Eq.~(\ref{EOM}) I will put off the discussion of his work to the
next subsection where the same formalism is also used.

Ankerhold \cite{Ankerhold2007} was looking for the correction
$S_1(x,v)$ in the form
$S_1(x,v)=\phi_0(x)+\sum_{n=1}^3v^n\phi_n(x)/n$ and got certain
conditions for the arbitrary functions $\phi_n(x)$'s from the
leading order solution of the Fokker-Planck equation accounting for
the noise with the nonzero third cumulant. When his ansatz is
plugged into Eq.~(\ref{HJlin}) one can easily find that the set of
equations obtained for $\phi_n(x)$'s is internally inconsistent
(suggesting that the truncation at the third power in $v$ in the
ansatz is insufficient) for a general $Q$ and potential
$V(x)$.\footnote{Of course, this is not too surprising in view of
the above-stated general properties of the non-equilibrium action,
see the comments below Eq.~(\ref{HJlin}).} There are several
exceptions when the inconsistencies are removed, namely for a
strictly harmonic potential $V(x)\propto x^2$ (this potential
doesn't exhibit a barrier, at least not a smooth one approximating
the tilted washboard potential) and in the limits either $Q\to 0$ or
$Q\to\infty$. This suggest that Ankerhold's solution
\cite{Ankerhold2007} could yield correctly the two limiting cases
$Q\to 0,\infty$ for the considered cubic approximation to the
potential. Indeed, in the limit $Q\to\infty$ his solution is equal
to mine for $s\to 1$ as already mentioned above.\footnote{This fact
is also not surprising since the ansatz used in my solution is just
a subset of his form of $S_1(x,v)$. The main difference is that I
used the ansatz only in the case where it does solve
Eq.~(\ref{HJlin}) and also the discussion of fixing the freedom in
the solution due to the homogeneous part etc. (present for any Q)
seems absent in his work.} The opposite limit $Q\to 0$ is simple
since that case is integrable for any strength of the third cumulant
and the linear response can be easily calculated analytically
yielding $R_{\Gamma}(Q\to 0)=\exp\left[2 D_2(s) 2e R^2 E_J^2
<\!\!<\!I_m^3\!>\!\!>/\hbar(k_B T_{\rm eff})^3\right] = \exp\left[2
D_2(s) Q^2 E_J^2 <\!\!<\!I_m^3\!>\!\!>/C I_0(k_B T_{\rm
eff})^3\right]$ with $D_2(s)=[(1+2s^2)\arccos s-3 s\sqrt{1-s^2}]/6
\approx 8\sqrt{2}/45(1-s)^{5/2}$ for $s\to 1$
\cite{Sukhorukov2007,Grabert2008}. This is identical to Ankerhold's
solution in the corresponding limit $Q\to 0,\,s\to 1$ (recall the
multiplicative correction factor of $2/3$ in the Erratum and
Ankerhold's definition of the bias-dependent quality factor of the
JJ $Q(s)=Q(1-s^2)^{1/4}$). Now, Ankerhold's solution can be
interpreted as a simple interpolation formula between the two
limiting cases reading $R_{\Gamma}(Q)=\exp\left[2 D(s, Q) E_J^2
<\!\!<\!I_m^3\!>\!\!>/C I_0(k_B T_{\rm eff})^3\right]$ with the
interpolating function
$D(s,Q)=D_2(s)Q^2/[1+Q^2D_2(s)/D_1(s)]=8/9\,Q^2(s)(1-s)^2/[Q^2(s)+5]$.
It turns out that Ankerhold's expression (Eq.~(13) of
Ref.~\cite{Ankerhold2007}) is a neat interpolation scheme between
the highly underdamped and overdamped junction limits. It certainly
provides a very efficient and quite precise interpolation formula
for a finite $Q$. Its detailed comparison with the numerically exact
solution will be shown in the next subsection.

\subsection{Numerical evaluation of the escape rate in a general situation}
\label{subsec22}

Now, we turn to the general case of the calculation of the rate
asymmetry for an arbitrary intensity of the non-equilibrium noise
acting on the junction. This is achieved by the numerical solution
of the effective dynamical system equations (\ref{EOM}). As already
mentioned the solution consists in finding a trajectory satisfying
the equations of motion (\ref{EOM}) and connecting in infinite time
(corresponding to the zero auxiliary energy $\mathcal{H}=0$) the two
fix-points $[x_{\rm min},0,0,0]$ and $[x_{\rm max},0,0,0]$ being the
(metastable) minimum of the potential and the top of the barrier,
respectively. There always exists a classical, ``relaxation"
solution corresponding to the dissipative but noise-free motion of
the effective particle from the barrier top down to the minimum.
This solution has $p(t)\equiv 0,\,y(t)\equiv 0$ and also the
associated action is zero. On the other hand we are interested in
the other, ``escape" solution connecting the two potential extrema
via trajectory with non-zero conjugated momenta $p(t),\,y(t)$. For
equilibrium, i.e.\ Gaussian, noise the two types of trajectories are
connected by (generalized) time-reversal which forms the basis of
the Onsager-Machlup theory and was used by Grabert
\cite{Grabert2008} for his linear response calculations. For general
non-equilibrium noise sources, however, the two trajectories are not
simply related and one has to calculate the escape trajectory
directly by solving the full system (\ref{EOM}).

This is exactly done here. The problem is formulated as a boundary
value problem (BVP) on an infinite time interval reflecting the
stationarity condition of the original escape problem. Obviously,
this makes the BVP rather tricky and one has to be cautious in its
solution. Once the solution $[x(t),v(t),p(t),y(t)]$ is found the
action difference between the two fix-points is calculated from the
definition as
\begin{eqnarray}\label{action}
    \Delta S  &\equiv  S(x_{\rm max},0)-S(x_{\rm
    min},0)=\int_{-\infty}^{\infty}dt\left[
    p(t)\dot{x}(t)+y(t)\dot{v}(t)-\mathcal{H}\big(x(t),v(t),p(t),y(t)\big)\right]\nonumber\\
     &= -\int_{-\infty}^{\infty}dt\left\{ Q^{-1}y(t)^2+\frac{\theta}{I_0\lambda}\left[\tilde{F}\left(-\frac{\lambda}{\theta}y(t)\right)
    -\tilde{F}'\left(-\frac{\lambda}{\theta}y(t)\right)\left(-\frac{\lambda}{\theta}y(t)\right)\right]\right\}.
\end{eqnarray}
The second line was obtained after using the explicit expressions
for $\dot{x}(t),\,\dot{v}(t)$ from Eq.~(\ref{EOM}) and $\mathcal{H}$
(\ref{Hamiltonian}) and the resulting action is thus expressed
solely via the conjugate momentum $y(t)$. The exponential part of
the rate asymmetry $\log(\Gamma_+/\Gamma_-)$ is then calculated as
the difference of the action for two opposite measured current
polarities.

Technically the BVP is formulated on a long, but finite time
interval estimated by the shooting solution used for the initial
guess, for details see below. At the ends of this time interval the
trajectory is assumed to be close enough to the respective fix-point
so that the linear approximation to the equations of motion
(\ref{EOM}) can be employed. The linearized system is then
characterized by the stability analysis which identifies
stable/unstable directions and corresponding eigenvalues. The
boundary conditions at the ends of the time interval are then
formulated with help of the respective linearized system in the
spirit of the study \cite{Beri2005}, Sec.~IV, i.e.\ the
2-dimensional unstable (stable) manifold around the minimum
(maximum) is identified and the solution is required to lie in them
which yields two boundary conditions at each fix-point. Solution of
the BVP is then sought for by a relaxation method with the `{\tt
bvp4c}' built-in solver in {\tt Matlab} \cite{Shampine2003}.

The solver needs a very good initial guess for the solution to
converge at all. When it does, it is very fast and efficient.
Without a good initial guess it usually does not converge,
especially for very large $Q$. Thus, the initial guess is the
crucial part of the whole solution process. For finding a good
initial guess I solve the BVP by the shooting method first (in line
with a general BVP strategy \cite{Shampine2003,Press1992}), i.e. I
look for a solution of Eq.~(\ref{EOM}) by solving an initial value
problem starting at the unstable manifold around the minimum and
searching for such an initial condition which evolves into the other
fix-point around the potential barrier. Finding a solution by the
shooting method solely is a rather difficult task due to involved
characteristic numerical instabilities around the target fix-point
\cite{Shampine2003,Kogan2008}. Indeed, the found solution is
typically not precise enough to be of any use for the evaluation of
the rates, however, it usually suffices as a good enough initial
guess to ensure the convergence of the relaxation method. Moreover,
the solution found by the shooting method gives a good estimate for
the time needed to join the two linearized manifolds around the
respective fix-points, a task which is not obvious how to accomplish
otherwise. After solving the problem at this fixed long time
interval I include corrections due to the rests of the infinite time
interval at the beginning and the end, respectively, by analytically
evaluating the associated quadratic action exactly analogously to
the method of Ref.~\cite{Beri2005}, Sec. V. These corrections,
although relatively small, are necessary within the precision
required by the problem. The last technical detail concerns the
handling of the (non)uniqueness of the solution. Since the system
(\ref{EOM}) is autonomous it has an infinite number of solutions
related by a simple time-translation, i.e. if
$[x(t),v(t),p(t),y(t)]$ is a solution then for an arbitrary $\tau$
shift in time $[x(t+\tau),v(t+\tau),p(t+\tau),y(t+\tau)]$ is also a
solution. This liberty may confuse the relaxation solver and, thus,
it is advisory to fix the solution by explicit breaking of the time
invariance \cite{Shampine2003}. I achieved this ``locally" by fixing
the phase of the oscillatory solution at the unstable manifold
around the potential minimum both in the boundary conditions as well
as in the initial value problem. Time shift by the period of the
local oscillations remains to be a symmetry operation but the
continuous symmetry with an arbitrary shift is this way broken. This
trick does help to stabilize the solution, nevertheless, despite of
all the tricks used the numerical implementation is still not
absolutely stable and one can occasionally run into problems of
non-convergence, especially with increasing value of $Q$. This is
not so surprising since for large $Q$ the time span needed to
connect the two linearized neighborhoods of the fix-points becomes
longer and the solution in between exhibits ever increasing number
of oscillations.

Further, due to the linear regime in which the problem is to be
solved to describe current experiments the effect of non-Gaussian
noise leads only to very small asymmetry effects and this requires
rather high precision of the calculations which stretches the used
method to its limits. Further improvement of the numerical method is
thus an interesting and urging open issue in the solution of the
present problem. During the {\tt UPoN2008} conference I became aware
of the work by the Lancaster group
\cite{Beri2005,LUCHINSKY2002,Khovanov2006} which has apparently
developed a toolbox of methods for tackling even far more difficult
problems of escape. While I developed some of the techniques they
use independently, there seem to be some more left to explore. It
will be interesting to see whether those techniques, such as the
action plot concept \cite{Beri2005}, will be successful in improving
the present numerics. My naive fast implementation attempts have
failed thus far. The main difference of the present problem from
most established techniques, including the Lancaster group's ones,
seems to lie in the presence of non-Gaussian noise and the task to
actually employ and characterize its effects on the escape
characteristics. It's unclear at this point how seriously this fact
influences the feasibility and/or performance of those techniques
developed predominantly for Gaussian problems.

\begin{figure}
  \centering
  \includegraphics[width=0.9\textwidth]{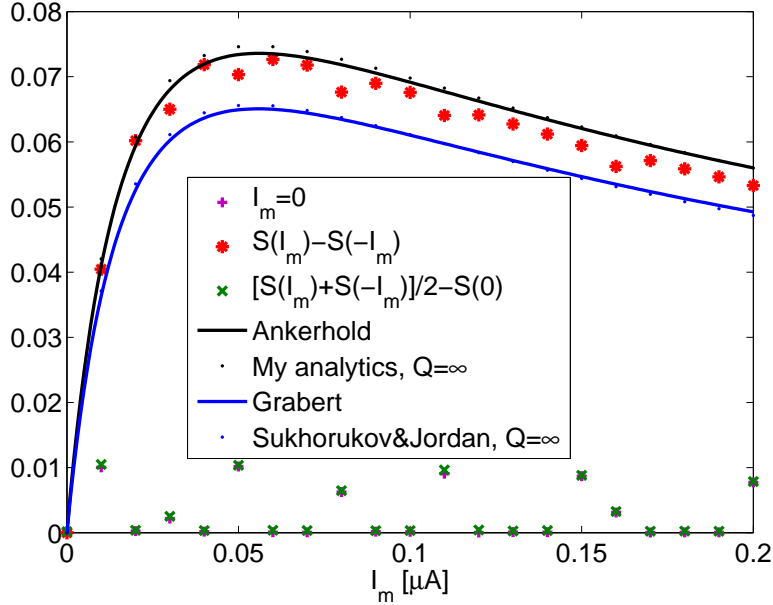}
  \caption{The logarithm of the escape rate asymmetry for temperature $T=20$~mK and quality factor $Q=22$
  corresponding to the Saclay experiment \cite{Huard2007}. For detailed explanation of various quantities
  and the values of other parameters see the main text.}
  \label{fig1}
\end{figure}

\begin{figure}
  \centering
  \includegraphics[width=0.9\textwidth]{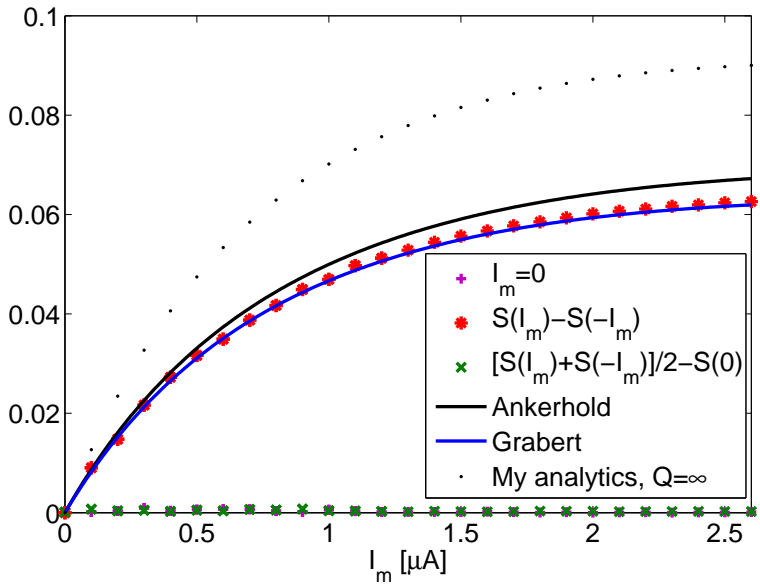}
  \caption{The logarithm of the escape rate asymmetry for temperature $T=200$~mK and quality factor $Q=4$.
  For detailed explanation of various quantities and the values of other parameters, which are
  the same as in Fig.~\ref{fig1} and correspond to the Saclay experiment \cite{Huard2007}, see the main text.}
  \label{fig2}
\end{figure}

For this moment I can only present numerical results obtained with
the ``not-yet-quite perfect" BVP method described above. The results
are shown in Figs.~\ref{fig1}, \ref{fig2} for parameters motivated
by the Saclay experiment by B.~Huard {\em et al.} \cite{Huard2007}.
Their junction was characterized by the critical current $I_0\doteq
0.48~\mu$A equivalent to the Josephson energy $E_J/k_B\doteq
11.4$~K, quality factor $Q\doteq 22$, and the dimensionless ``kick"
parameter $\lambda\doteq 0.002$. The corresponding plasma frequency
of the unbiased junction was $\omega_{p0}\approx 1$~GHz. These
experimental parameters correspond exactly to Fig.~\ref{fig1} while
in Fig.~\ref{fig2} I just modified the value of the quality factor
$Q=4$ ``by hand" (corresponding to changing the value of the
resistance $R$ while keeping the other parameters constant) to
explore a more intermediate regime of $Q$ and see the performance of
different theories also there, not only in the high $Q\to\infty$
limit as in Fig.~\ref{fig1}. The experiments were performed for
temperatures in the range 20--530~mK. This is reflected by the
lowest temperature of $T=20$~mK used in Fig.~\ref{fig1} while
Fig.~\ref{fig2} presents results for an intermediate temperature of
$T=200$~mK. Compared with the Josephson coupling energy $E_J$ both
the reservoir as well as effective temperatures are small, on the
order of few per-cents, which justifies the usage of the above
developed theory of the non-equilibrium action valid for the weak
noise only.

The fast inspection of both figures reveals that the logarithm of
the asymmetry $\log(\Gamma_+/\Gamma_-)$ presented there is generally
a small quantity with a typical magnitude of about 10\%. This is
consistent with the usage of the linear response theories (in the
third cumulant) employed in previous studies
\cite{Sukhorukov2007,Grabert2008,Ankerhold2007}. The curves are not,
however, linear functions of the measured current $I_m$ due to the
fact that the current contributes also to the effective temperature
(and it is an important contribution) which enters the formula for
the asymmetry linear in the third cumulant (proportional to $I_m$ as
well for the tunnel junction). Moreover, there is another source of
nonlinearity in the curve, namely the fact that experimentalists for
convenience measure in the range of roughly constant mean escape
rate on the order of $\approx 30$~kHz. Thus, for each value of the
measured current $I_m$ the (dimensionless) bias current $s$ is
adjusted in such a way that the mean rate stays constant close to
that value. For the present junction with $\omega_{p0}\approx 1$ GHz
this implies fixing the dimensionless barrier height to a value of
roughly 10.4. In other words, for every $I_m$ the value of $s(I_m)$
is determined from the equation $\Delta U(s(I_m))/k_BT_{\rm
eff}(I_m)\approx-\log(3\cdot10^{-5})\doteq 10.4$ with $\Delta U(s)=2
E_J(\sqrt{1-s^2}-s\arccos s)\approx 4\sqrt{2}E_J(1-s)^{3/2}/3$ for
$s\to 1$ being the barrier height of the tilted washboard potential
and $T_{\rm eff}(I_m)=T+eRI_m/2k_B$ the effective temperature.

Let us first discuss Fig.~\ref{fig1} with high $Q=22$. This value of
the quality factor was nearly at the edge of stability of my BVP
numerics described above. The diagnostic quantities are shown in the
plot to exemplify the precision achieved in the calculations. I plot
the quantity $S(0)-S_{\rm anal}(0)$ to assess the overall precision
of the calculation. By $S(0)$ I denote the action calculated
numerically for zero measured current $I_m=0$. This value is known
analytically and equals the already discussed experimental value of
$\approx 10.4$. This quantity, in the plot labeled `$I_m=0$' is
shown by pluses which are essentially overlaid by crosses. The
crosses show the quantity $[S(I_m)+S(-I_m)]/2-S_{\rm anal}(0)$ which
probes the linearity of the calculated action in the third cumulant.
If in the linear regime, the two polarities contain opposite
contributions from the third cumulant which cancel in the sum and
the subtracted action for zero third cumulant should nullify this
quantity. In the second Fig.~\ref{fig2} with $Q=4$ this is indeed
the case due to more stable numerics but one can see that those two
control quantities are not strictly zero for the high-Q case. Their
overlap, however, actually confirms the linear response regime. The
deviation from zero of $[S(I_m)+S(-I_m)]/2-S_{\rm anal}(0)$ are
solely due to the imprecision of the mean action without any
influence of the third cumulant. This is further confirmed by the
essentially regular behavior of the asymmetry $S(I_m)-S(-I_m)$.
Moreover, it should be stressed that each point presents an
independent calculation. Thus, the values of $I_m$ where the control
quantities are zero as expected should be trustworthy regardless of
the fact that the next value of $I_m$ may be calculated with
insufficient precision. Moreover, the overall precision even in the
$Q=22$ case is not catastrophically bad although it does not allow a
fully reliable comparison with the concurrent theories.

The asymmetry $S(I_m)-S(-I_m)$ in Fig.~\ref{fig1} is compared with
four different theories grouped into two sets (within the set they
are virtually equal in the $Q\to\infty$ limit). It is (generalized
numerical) evaluation $\grave{a}$ la Grabert \cite{Grabert2008}
together with the result by Sukhorukov and Jordan
\cite{Sukhorukov2007} both of which predict basically
$S(I_m)-S(-I_m)\propto 0.79(1-s)^2$ while the other set is my
Eq.~(\ref{D1}) and Ankerhold's \cite{Ankerhold2007} result
$S(I_m)-S(-I_m)\propto 8/9(1-s)^2\doteq 0.89(1-s)^2$. While the
difference in the predictions is only on the order of 10\% and,
thus, most likely irrelevant for experiments, it is relevant from a
purely conceptual point of view which one is actually correct since
it should help with the identification of possible misconceptions
hidden in the failed approach(es). From the data presented in
Fig.~\ref{fig1} it is clear that the more promising set is the
Ankerhold-Novotn\'{y} one. Despite of the scatter in the data, there
are reliable points (where the control quantities turn into zero)
which are closer to the 8/9-curve. The numerical calculation didn't
use any linear perturbation theory or any approximation at all. The
data are sheer results of the numerical evaluation of the BVP for
general values of the parameters. The discrepancy of the data with
the theoretical predictions may be caused by the finite, although
rather high, value of $Q$. This can account for the difference
between the numerics and 8/9-curve, however, it is inconsistent with
Grabert's theory which predicts monotonic increase of the asymmetry
with increasing $Q$, see Fig. 4 in Ref.~\cite{Grabert2008}.
Moreover, ``Grabert's curve" was calculated for $Q=22$ even though
it hardly deviates from its limiting $Q=\infty$ counterpart by SJ.
Thus, the only salvation for the two theories
\cite{Sukhorukov2007,Grabert2008} could come from the numerics being
wrong which is in principle possible but does not seem too plausible
at this point.

If we now turn to the other Fig.~\ref{fig2} we see at the first
place much better precision of the numerics as revealed by the
control quantities being zero. The numerical data are again compared
with Grabert's and Ankerhold's theories which provide alternatives
for finite $Q$. I also show a curve for $Q=\infty$ to demonstrate
significant deviations of the results for still relatively high
$Q=4$ from the infinite-$Q$ limit. This should be remembered when
interpreting experimental data of, e.g., the Helsinki group
\cite{Timofeev2007,Peltonen2007} with $Q\approx 2.5$ via $Q=\infty$
theories. Ankerhold's theory (Eq.~(13) of Ref.~\cite{Ankerhold2007}
with the 2/3-correction from the Erratum) is off the numerical data
as well as Grabert's result thus clearly demonstrating only the
interpolating status of this theory. The discrepancy is, however,
rather small and, therefore, Ankerhold's formula seems to be a very
cheap and efficient analytical interpolation scheme for an arbitrary
$Q$.

Grabert's result on the other hand lies exactly on top of the
numerical data in stark contrast to its apparent failure for the
high-$Q$ case in Fig.~\ref{fig1}. This is somewhat mysterious
behavior which certainly deserves better understanding. What could
go wrong in Grabert's reasoning? I have no clear answer to that,
however, I do have a conjecture where there could be a problem
hidden. Of course, I am fully aware that the problem could in fact
be also in my numerics for $Q=22$ although its correspondence with
my analytics represented by Eq.~(\ref{D1}) is encouraging and not
quite typical for bug-plagued numerics. Grabert's approach uses a
straightforward perturbation theory at the level of trajectories
connecting the fix-points. He argues that within the linear response
in the third cumulant the equilibrium (unperturbed) solution is
enough for evaluating the correction to the action. In more detail,
provided the auxiliary Hamiltonian is split into equilibrium part
and non-Gaussian perturbation $\mathcal{H}=\mathcal{H}_{\rm
equil}+\mathcal{H}_3$ the correction to the action reads $\Delta
S_3=-\int_{-\infty}^{\infty}dt \mathcal{H}_3(y_{\rm equil}(t))$
(compare with his Eq.~(77) in Ref.~\cite{Grabert2008}). This is
analogous to the standard first order perturbation theory in quantum
mechanics, the correction to energy is just the mean value of the
Hamiltonian in the unperturbed state. However, one should recall
that this formula is only applicable if the unperturbed state is
non-degenerate. It's not obvious what is the analogous condition for
classical trajectories, nevertheless, one may expect certain
subtleties involved due to several conditions specific for the
current problem. First of all, the BVP is formulated on an infinite
time interval, there exists the continuous time shift symmetry, and
the unstable/stable manifolds around the respective fix-points are
two-dimensional (could this be the ``degeneracy"?). The above
formula for $\Delta S_3$ can be easily derived for finite time
interval with fixed boundary conditions, however, can't the infinite
time interval bring about omitted surface terms? I am quite sure
these questions can be successfully handled by dynamical system
theory experts.

\section{Experiment-related issues}
\label{Sec3}

To this date (end of June 2008) there are two publicly available
experimental results by the Helsinki group
\cite{Timofeev2007,Peltonen2007} and by the Saclay group
\cite{Huard2007}. The Helsinki experiment finds the asymmetry curve
as a function of the measured current through the tunnel junction
which has its shape in qualitative agreement with all previously
mentioned theories (the $\sim10\%$ difference between different
theories is undetectable at the level of precision of the
experiment). However, the quantitative comparison with, e.g.,
Ankerhold's theory shows discrepancy on the order of $\sim 10$ (see
the comparison in Ref.~\cite{Peltonen2007}, recall the correction
factor 2/3 missing in that reference and further account for finite
$Q=2.5$ contributing another factor of ~1/2). I haven't discussed
the theory used by the Helsinki group for fitting the experiment
since it's conceptually different from all the other discussed
theories and I consider it to be semi-phenomenological with the
prefactor (calculated in other theories) being adjusted to the
experimental outcome, thus lacking a real predictive power. The
other experiment by the Saclay group has been identified as most
likely faulty due to a leak in the measurement circuit which
prohibited the reliable determination of the bias current. Such an
effect largely overshadows any asymmetry due to the third cumulant
and, thus, no quantitatively reliable data are available from this
experiment.

Regardless of this unsatisfactory status we may consider possible
problems which are likely to be encountered, and maybe have already
been encountered in the Helsinki experiment, when trying to compare
the experimental outcome with theoretical predictions. The first
issue is the one of the actual relevance of the exponential part of
the rate asymmetry. Clearly, the experiment measures the rate
asymmetry, not a theoretical concept of its exponential part. The
rationale behind the dominance of the exponential part of the rate
unfortunately doesn't necessarily carry over to the {\em rate
asymmetry}, especially in the {\em linear regime}. The standard
argument behind the dominance of the exponential part of (thermal)
rates is that the large dimensionless barrier entering the exponent
simply dominates the whole expression; moreover, the noise intensity
(temperature) enters only the exponential part via the Boltzmann
factor while the prefactor (attempt frequency) is
temperature-independent. Now consider a weak noise with the third
cumulant nonzero. This weak noise will supposedly weakly modify the
rate. This will in general happen both through the exponent and the
prefactor. In the linear response regime in the third cumulant the
correction in the exponent can be safely expanded and the resulting
linear correction will add to the linear correction stemming from
the prefactor. At this stage there is no a priori difference between
these two contributions. Of course, in practice one of them
(presumably the prefactor part) can still be negligible. What are
the prospects for this to happen? We have seen that in the realistic
setup studied in the previous section the asymmetry due to the
exponential part of the rate reaches values on the order of
$\sim10\%$ at maximum. The expected correction due to the prefactor
is of the form $k_BT_{\rm eff}/E_J\cdot I_m/I_0$. The first factor,
dimensionless temperature, is of the order of $\sim 1\%$ while the
other factor, dimensionless measured current, is of the order of
$\sim 1$. Thus, in total, we have an effect of the order of $\sim
1\%$ which can be, depending on the actual numerical prefactor,
comparable to the exponential part. This somewhat pessimistic
scenario can be further supported both by the discrepancy found in
the Helsinki experiment as well as by the mismatch between
Ankerhold's theory and direct stochastic simulations performed in
connection with the Saclay experiment in Ref.~\cite{Huard2007} (see
their Fig.~7a) where a multiplicative factor of 2 difference was
found for the dimensionless barrier height $\sim 6$. While this
value lies at the border of reliability of the WKB approach and
corrections for larger barriers (especially the experimentally
relevant one 10.4) may be expected, they are not expected to be of
order of 100\%. Therefore, it seems that the asymmetry stemming from
the prefactor may be relevant for experiments. This is a rather bad
news for theoreticians since the calculation of the prefactor for
non-equilibrium rates is an involved task, see the discussion in
Ref.~\cite{Hanggi1990} and references therein.

So we finally come to the question whether one can achieve a
nonlinear regime with underdamped JJs. The problem is apparently in
the fact that the effective temperature raises with the measured
current in such a way that it simply dominates the escape mechanism
and corrections due to higher order cumulants are just negligible.
This is clearly reflected in the plot in Fig.~\ref{fig1} where the
originally growing (with $I_m$) curve for small $I_m$ eventually
bends downwards again for larger $I_m$. While the first part is
governed by the third cumulant growing with $I_m$, the declining
part corresponds to the case when the contribution of $I_m$ to the
effective temperature beats the raising third cumulant. The same
effect would be seen in Fig.~\ref{fig2} for larger values of $I_m$.
This behavior could be diminished by weakening the effect of the
measured current on the effective temperature, see the expression
for the effective temperature. This should be achieved by decreasing
the value of the effective shunt resistance $R$. This, in turn,
would imply decreasing quality factor which seems experimentally
unacceptable beyond the point when the switching ceases to exist and
only phase diffusion is present. The quality factor thus should be
maintained at a reasonably high value which can be achieved by
increasing capacitance $C$. That in turn will decrease the third
cumulant contribution via the formula preceding Eq.~(\ref{D1}). At
this point the problem turns into a bad joke. There may be, however,
a parameter window where a subtle compromise can be achieved. This
should be seriously considered by carefully examining different
parameter dependencies and testing experimentally acceptable
numbers.

\section{Conclusions}
\label{Concl}

In this work I have reviewed in detail the status of the problem of
the measurement of the Full Counting Statistics by the switching
dynamics of an underdamped Josephson junction. I have presented a
general theory for the weak noise based on the WKB-like
approximation and calculated the rate asymmetry due to a weak third
cumulant analytically in the limit of very high quality factor of
the junction. This calculation has been critically compared to other
theories and their possible shortcomings have been identified and
pointed out. Further, I have developed a numerical scheme for
solving the boundary value problem determining the exponential part
of the non-equilibrium escape rate under general circumstances,
i.e.~beyond the linear perturbation theory. Using this scheme I have
calculated the exponential part of the rate asymmetry for
experimentally relevant set of parameters and compared the findings
with various linear theories. Again, this helped with the
identification of the status of concurrent theories. Eventually, I
have briefly discussed issues related to the interpretation of
present and future experiments, in particular the question of the
relevance of the rate exponential prefactor for the rate asymmetry
and the feasibility of achieving the nonlinear regime.

There are plenty of unsolved problems and open issues within this
field of research. Starting with those more particular and
technical, it would be rewarding to fully clarify the status of
concurrent theories, in particular that by Grabert which performs
amazingly well for intermediate range of $Q$'s while it seems to
fail for large values of $Q$. Although the discrepancy is not too
large, Grabert's theory is supposed to work in that regime as well
and, thus, the discrepancy raises serious questions about its very
foundation. On the other hand, it would be very helpful to further
develop and fully stabilize (if possible) my numerical scheme used
for the solution of the BVP. If this attempt were successful the
numerical code could be used for interpreting future experiments
routinely since it is very fast and efficient as long as it
converges which unfortunately occasionally doesn't happen. On a more
general level, it should be further studied what is the effect of
nominally subleading terms in the rate on the rate asymmetry. It
appears that the conventional arguments for the dominance of the
exponential part of the rate may not be applicable to the rate
asymmetry, especially in the linear regime. And last but not least,
a most important question whether one can actually use underdamped
Josephson junctions for the measurement of the whole FCS and not
only the third cumulant in the linear regime is still open and
waiting for final answer which, if affirmative, could bring the
field of the FCS to new milestones.

\section*{Acknowledgments} I am grateful to P.~H\"{a}nggi, I.
Khovanov, K. Neto\v{c}n\'{y}, and P. Talkner for useful and
stimulating discussions, continuous encouragement, and for pointing
out relevant references. I acknowledge the support by the grant
202/07/J051 of the Czech Science Foundation. This work is also a
part of the research plan MSM 0021620834 financed by the Ministry of
Education of the Czech Republic.

{\small {\em Post-acceptance note.} After the acceptance of this
manuscript with minor corrections I became aware of a comment {\tt
arXiv:0807.2675} by Sukhorukov and Jordan. Although at this point I
am unable to decide whether that comment really settles the above
mentioned discrepancy between our theories, I recommend the
interested reader to check out their paper for an independent point
of view on the issue.}
\newline

\end{document}